\documentclass[twocolumn,showpacs,aps,amsmath,amssymb,citeautoscript,longbibliography,floatfix,superscriptaddress,reprint]{revtex4-2}

\usepackage{mathtools} 
\usepackage{amsmath, amsthm, amsbsy, amssymb}
\usepackage[caption=false]{subfig}
\usepackage[colorlinks, citecolor=blue, linkcolor=blue, urlcolor=blue]{hyperref}
\newcommand{\E}	[1]		        {e^{#1}}
\newcommand{\D}                 {\mathrm{d}}
\newcommand{\abs} [1]           {\left| #1\right|}
\newcommand{\bk} [1]           {\left( #1 \right)}
\newcommand{\bkk} [1]           {\left[ #1 \right]}

\begin{document}

\title{Directional electron--filtering at a superconductor--semiconductor interface}

\newcommand{\itpa}{Institut f\"ur Theoretische Physik und Astrophysik,
	Universit\"{a}t W\"{u}rzburg, Am Hubland, D-97074 W\"{u}rzburg, Germany}
\newcommand{\physinst}{Physikalisches Institut, Universit\"{a}t W\"{u}rzburg, Am Hubland, D-97074 W\"{u}rzburg, Germany}
\newcommand{\delft}{Kavli Institute of NanoScience, Faculty of Applied Sciences, Delft University of Technology, Lorentzweg 1, 2628 CJ Delft, The Netherlands}
\newcommand{\ctqmat}{W\"urzburg-Dresden Cluster of Excellence ct.qmat, Germany}

\affiliation{\itpa}
\affiliation{\delft}
\affiliation{\physinst}
\affiliation{\ctqmat}

\author{Daniel Breunig}
\affiliation{\itpa}

\author{Song-Bo Zhang}
\affiliation{\itpa}

\author{Bj\"orn Trauzettel}
\affiliation{\itpa}
\affiliation{\ctqmat}

\author{T. M. Klapwijk}
\affiliation{\delft}
\affiliation{\physinst}
\affiliation{\ctqmat}

\date{\today}

\begin{abstract}
We  evaluate the microscopically relevant parameters for electrical transport of hybrid super\-conductor--semiconductor interfaces. In contrast to the commonly used geometrically constricted metallic systems, we focus on materials with dissimilar electronic properties like low--carrier density semiconductors combined with superconductors, without imposing geometric confinement. We find an intrinsic mode--selectivity, a directional momentum-filter, due to the differences in electronic band structure, which creates a separation of electron reservoirs each at the opposite sides of the semiconductor, while at the same time selecting modes propagating almost perpendicular to the interface. The electronic separation coexists with a transport current dominated by Andreev reflection and low elastic back-scattering, both dependent on the gate-controllable electronic properties of the semiconductor.
\end{abstract}
\maketitle


\section{Introduction}
\begin{figure}[ht]
	\renewcommand{\figurename}{Fig.}
	\centering
	\includegraphics[width=.99\linewidth]{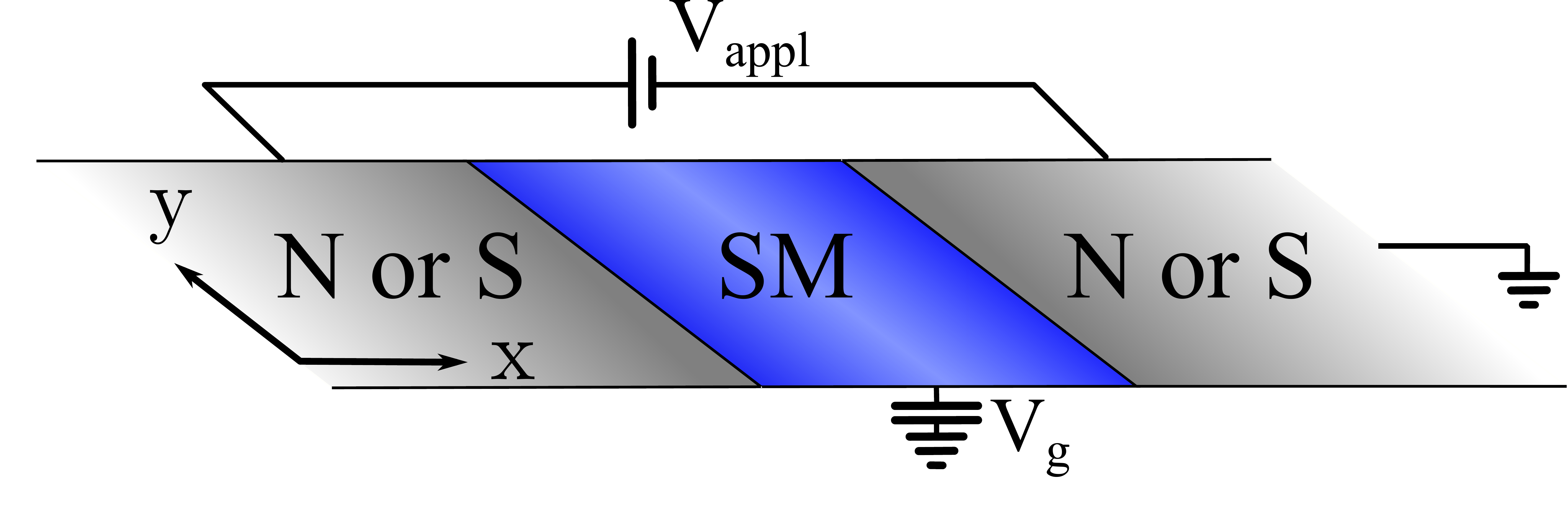}
	\caption{Sketch of a two-dimensional interface-junction considered in this work.  A semiconductor (SM, blue) is brought in contact to a normal metal (N, gray) or an s-wave superconductor (S, gray). We assume two interfaces in $x$--direction, while the $y$--direction remains translational invariant.}
	\label{fig:1}
\end{figure}

The recent interest in the interaction of topological materials and conventional superconductors raises questions about the electronic properties of their interfaces. In general, in the analysis of quantum transport properties of one-dimensional edge channels, a distinction is made between two-point and four-point measurements \cite{Buettiker1988,Stone1988,Beenakker1991}. A Josephson-junction is intrinsically a 2-terminal device in which the same contact is used for source or drain, as well as for the voltage terminals. Although quantum-transport analysis works very well for constriction-type metallic Josephson-junctions consisting of one and the same material \cite{BeenHout1991,Nazarov2009},  an application to Josephson-junctions consisting of dissimilar materials with wide planar interfaces (Fig.\ref{fig:1}) is often implicitly assumed, but not obviously justified. We will demonstrate that hybrid Josephson-junctions deserve an explicit analysis, based on the different electronic properties \cite{Blonder83, Buettiker1985,Landauer1987,Landauer1995}. The useful feature of semiconductors, that  the electronic properties can be changed by changing the Fermi level with a gate, also affects the electronic transport-properties of the interface. These interface properties play a crucial role in mediating the macroscopic phase coherence between the two superconductors across the semiconductor, not only quantitatively, but also conceptually. This aspect becomes particularly urgent when studying the \textit{voltage-carrying state} of a ballistic hybrid Josephson-junction, but it is also important for the zero-voltage state because of its effect on the boundary conditions for the Andreev bound states \cite{Radovic2009}.
\par 
Our analysis starts with the experimental observation of a so-called \textit{excess current} in Josephson-junctions (like Fig. \ref{fig:1}) based on HgTe-heterostructures using superconductors like Al, Nb, and MoRe \cite{Oostinga2013,Wiedenmann2015,Bocquillon2016}. The devices consist of large cross-sectional areas of a few micron by 10 to 80 nanometer. Different arrangements are possible such as planar electrodes on the active surface of the HgTe-layer, in which it is assumed that superconductivity is induced by the \textit{proximity-effect}. More recently, mesa-structures are used with superconducting contacts made on the sides of the thin HgTe-layer \cite{Bendias2018} (like also used for graphene \cite{Calado2015} and, very recently, for InSb quantum wells \cite{Ke2019}). In all cases a characteristic excess current is observed for applied voltages higher than the superconducting energy-gap. This feature is since Blonder \textit{et al.} \cite{Blonder1982} understood as due to enhanced charge transport by Andreev-reflection for a range of electron-energies, which match the energy-gap. The principle can be understood by studying the quantum-mechanics of a single interface between a superconductor and a normal material. In order to make these interface-properties observable in a conductance experiment Blonder \textit{et al.} have chosen a suitable geometry, compatible with the experiments. It positions this transmissive interface in an orifice in an otherwise opaque screen, which separates the electron systems of the two materials.  This separation allows the assignment of different chemical potentials to the two reservoirs \cite{ImryLandauer1999} with the occupation in one reservoir characterized by $f_0(E)$ and the other one by $f_0(E+eV)$ with $f_0$ the Fermi-Dirac distribution. For a given applied voltage, the conductance through the orifice is calculated as the difference between the flux of right-movers and left-movers, like it has become customary in studying quantum point contacts. In this framework, an orifice has resistance, even in the absence of backscattering, known as the Sharvin-resistance. It is assumed that the diameter of the orifice is smaller than the elastic mean free path, which is feasible for metallic point contacts and also for tunnel-junctions with a pinhole, but it is definitely not in agreement with the dimensions used in the HgTe-samples \cite{Oostinga2013,Wiedenmann2015,Bocquillon2016,Bendias2018}. In these electronic transport experiments an excess current is found, which indicates very weak elastic back-scattering, \textit{i.e.} a high transmission-coefficient, which scales with the area. The conceptual problem is therefore to understand the experimental observation of a conductance of a geometrically large planar interface with excellent Andreev-reflection properties, including a low value of elastic back-scattering, while missing in the experimental setup an obvious mechanism through which the highly transmissive interface could also provide a separation of the two electronic reservoirs, each at a different chemical potential determined by the applied voltage.


\section{Fermi surface mismatch}
\begin{figure}[htpb]
	\renewcommand{\figurename}{Fig.}
	\centering
	\includegraphics[width=.99\linewidth]{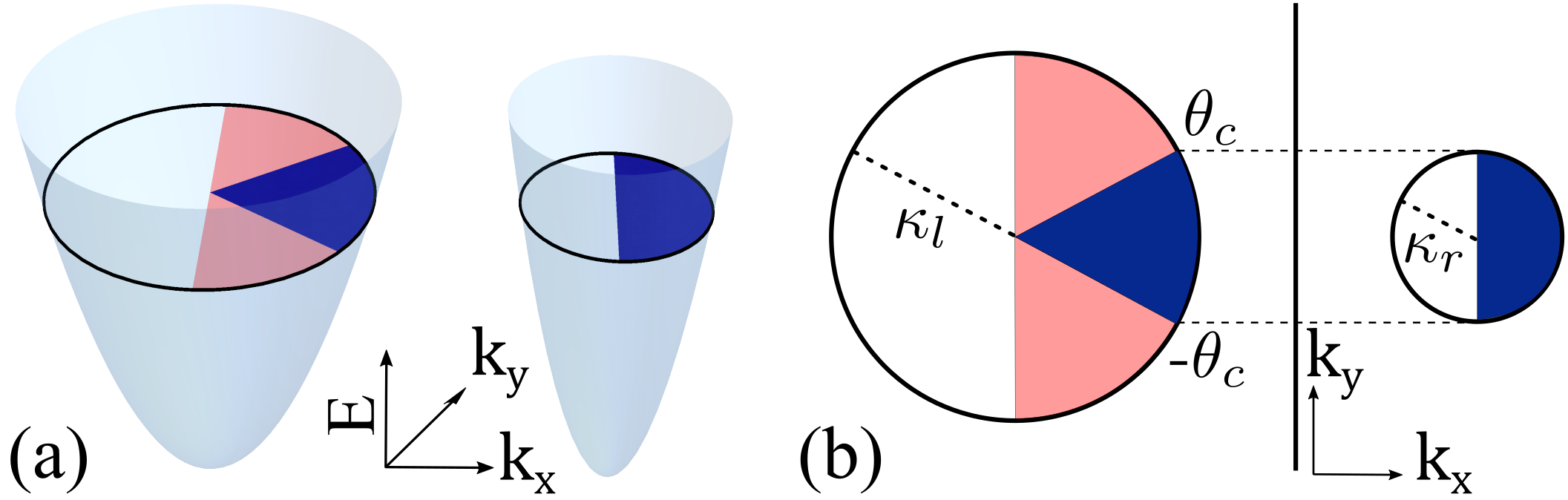}
	\caption{(a) 3D and (b) top view illustration of the FSM. At any energy, electron states lie on circles with different radii $\kappa_{l}$ and $\kappa_{r}$ in $k$--space. Only a fraction of the states (blue sectors) can contribute to transmission, the rest (red sectors) is reflected. $\theta_c$ represents the critical angle of incidence. }
	\label{fig:2}
\end{figure}
The key concept in this work is Fermi surface mismatch (FSM), which we illustrate for an NN'--junction in Fig. \ref{fig:2}. Assuming different Fermi-momenta, for instance, due to different effective masses at the left ($L$) and right ($R$) part of the interfaces, the available electron states lie on circles in momentum space with generally different radii. For clarity, we focus here on the two--dimensional (2D) case and isotropic Fermi--surfaces. Because of the different size of the Fermi-surfaces, there are modes in $L$ with no corresponding modes in $R$, \textit{i.e.} electrons with an angle of incidence larger than a critical angle $\theta_c$ cannot be transmitted across the interface, but are reflected back. Consequently, they do not contribute to the conductance across the junction, in contrast to the modes with an angle of incidence $|\theta |<\theta_c$ (while possibly being limited by an additional scattering  potential). In many real systems, a quasi-2D semiconductor is coupled to 3D metallic reservoirs. This implies that many modes of the 3D metals (corresponding to a mode index that parameterizes the direction perpendicular to the quasi--2D system) are not transmitted into the semiconductor. If two regions are weakly coupled in space, it is possible to define separate thermal equilibria for each of them. We argue here that FSM is a feasible mechanism to decouple (and therefore only weakly couple) two regions in space with a larger Fermi surface by one region in space with a smaller Fermi surface, \textit{cf}. Fig. \ref{fig:1}.
\par
FSM is not a new concept \cite{Kashiwaya1996, Linder2008, Cayssol2008, Breunig2018}. Our main message, however, is its effect on the observable conductance and, more importantly, its meaning for the proper definition of local thermal equilibria and thus the applicability of contemporary transport theories. 
 We demonstrate below that in analogy to the well-known Landauer conductance in a geometrically confined geometry of $G={2e^2}/h\sum |t_{n,m}|^2$, the system has a conductance per unit length given by  
\begin{align}
	\label{eq:1} 
	G_{0e} = \frac{2e^2}{h} \kappa_l \int\limits_{-\theta_c}^{\theta_c}\D \theta_l\cos\theta_l\; T\bk{\theta_l}
\end{align}
with $\theta$ the angle of incidence of electrons approaching the interface, $\theta_c$ a critical angle determined by the electronic mismatch between the reservoirs and the semiconductor, and $T\bk{\theta_l}$ the transmission-coefficient for the incident wave \footnotemark[3]. The conductance is tunable by changing the Fermi-level of the semiconductor, which changes $\theta_c$ and  $T\bk{\theta_l}$. The system can be viewed as providing a confinement in momentum space rather than in real space as anticipated by B\"uttiker \cite{Buettiker1985} and  Landauer \cite{Landauer1987}, assuming the absence of any limiting contribution to the conductance by elastic or inelastic scattering. It is analogous to the \textit{geometric} Sharvin--resistance.  In Eq. \eqref{eq:1}, it is labeled $G_{0e}$ to indicate its source in the electronic mismatch between the two material-systems, and could be called an \textit{electronic} Sharvin--resistance.  
\par
We analyze the quantum-mechanical properties of a single interface from the metallic superconductor into the semiconductor. Based on this analysis we will argue that, to make the interface properties observable in a conductance experiment \cite{Oostinga2013,Wiedenmann2015,Bocquillon2016}, one needs two interfaces. Since the excess current is a property observable at voltages higher than the superconducting energy gap, the crucially relevant energy-conditions are then equivalent to those of one interface, the semiconductor and the superconductor. The experimental data at lower voltages will require, in addition to the quantum-mechanical interface properties, the determination of a non-equilibrium distribution function for the occupation of the states in the semiconductor in the spirit of the SINIS-model discussed by Octavio \textit{et al.} \cite{Octavio1983,Flensberg1988}, but including the directionality.
\par

\section{Model}
For the S contact, we assume intrinsic s--wave pairing and use the Bogoliubov--de Gennes (BdG) formalism \cite{deGennes}. Introducing the Nambu spinor $(\hat c,\;\hat c^\dagger)^T$, where $\hat c^\dagger$ is the electron creation operator, we define the BdG Hamiltonian as
\begin{align}
	\label{eq:2}
	\mathcal H_{BdG}(x)=[\mathcal H_{0}(x)+H\delta(x)]\tau_z+ \Delta(x)\tau_x,
\end{align}
where $\tau_{x,z}$ are Pauli matrices and $\mathcal H_0$ the normal--state Hamiltonian
\begin{align}
	\label{eq:3}
	\mathcal H_0(x)&=\hat k_x\frac{\hbar^2}{2m(x)}\hat k_x+\frac{\hbar^2k_y^2}{2m(x)}-\mu(x)
\end{align}
with $\Delta(x)$ the position--dependent order parameter in the S region and $H$ a localized repulsive potential to model an interface with a conventional elastic scattering potential. The form of Eq. \eqref{eq:3} guarantees that the Hamiltonian is hermitian and well defined \cite{davies1998physics,horing2017quantum}, since the momentum operator in $x$--direction, $\hbar\hat k_x=-i\hbar\partial_x$, does not commute with a position--dependent effective mass $m(x)$. Here, $\hbar$ is the reduced Planck constant, $\mu(x)$ the electrochemical potential, related to the carrier density, and $k_y$ is the $y$--component of the wave vector, which parameterizes the transverse modes \cite{datta1995}. For simplicity, we assume quadratic dispersion relations in all parts. However, all our findings only depend quantitatively but not qualitatively on this choice. 
\par
The left and right domain, which are N and S, respectively, are described by $\Delta( x)=\Delta_0\Theta(x)$, with $\Theta\bk{\cdot}$ the Heaviside function. The effective masses are assumed to be constant in each part, but not equal, $m(x)=m_l\Theta(-x)+m_r\Theta(x)$. Furthermore, we assume a global electrochemical potential, $\mu(x)=\mu$, which is significantly larger than the excitation energy $E$ and the order parameter $\Delta_0$, i.e. $\mu\gg E,\Delta_0$ \footnotemark[1]. Under these assumptions (for detailed derivations, see App. \ref{app:A} and \ref{app:B}), the eigenstates can be written as
\begin{align}
	\label{eq:4}
	\psi^\pm_e(x)= \begin{pmatrix}1 \\0\end{pmatrix}\E{\pm i k_l x}, &&
	\psi^\pm_h(x)= \begin{pmatrix}0 \\1\end{pmatrix}\E{\mp i k_l x},
\end{align}
in $L$ and
\begin{align}
	\label{eq:5}
	\psi^\pm_{eq}(x)= \begin{pmatrix}u \\v\end{pmatrix}\E{\pm i k_{r} x}, &&
	\psi^\pm_{hq}(x)= \begin{pmatrix}v \\u\end{pmatrix}\E{\mp i k_{r} x},
\end{align}
in $R$, where the superscript $(\pm)$ indicates the group velocity with respect to the $x$--axis. The subscript distinguishes electrons ($e$), holes ($h$), electron--like ($eq$), and hole--like ($hq$)  quasiparticles. Defining $\theta_{l/r}=\arcsin\bk{k_y/\kappa_{l/r}}$, the wave numbers and the associated group velocities are given by $k_{l/r}=\kappa_{l/r}\cos \theta_{l/r}$ and 
\begin{align}
	\label{eq:6}
	v_l=\frac{\hbar k_l}{m_l}, &&
	v_r=\frac{\hbar k_r}{m_r}\bk{\abs{u}^2-\abs{v}^2},
\end{align}
respectively. Here, $\kappa_{l/r}=\sqrt{2m_{l/r}\mu}/\hbar$ are the Fermi wave numbers (\textit{cf.} Fig. \ref{fig:2}) and $u^2=1-v^2=\bk{1+\Omega/ E}/2$ with $\Omega=\sqrt{E^2-\Delta_0^2}$ the superconducting coherence factors. Using scattering theory, we calculate the probabilities for Andreev \cite{Andreev64} ($A$) and normal ($B$) reflection and thus the transmission coefficient for each incident electron as
\begin{align}
	\label{eq:7}
	T(E,\theta_l)=1+A(E,\theta_l)-B(E,\theta_l),
\end{align}
which is energy and mode--dependent.
\par

\section{Angle-dependent transmission}
 We start our analysis with an NN' junction, by setting $\Delta_0=0$ everywhere, while maintaining FSM. Several results and conclusions found in NS systems can be inferred from the normal state transmission, providing a convenient theoretical foundation for the subsequent analysis. To quantify FSM, we define the ratio of the Fermi wave numbers as $r\equiv\kappa_r/\kappa_l=\sin\theta_c$, with $0< r\le 1$. The results for the complementary regime, $r>1$, are essentially the same, corresponding to the mirror image of  Fig. \ref{fig:2}(b).

\begin{figure}[htpb]
	\renewcommand{\figurename}{Fig.}
	\centering
	\includegraphics[width=.99\linewidth]{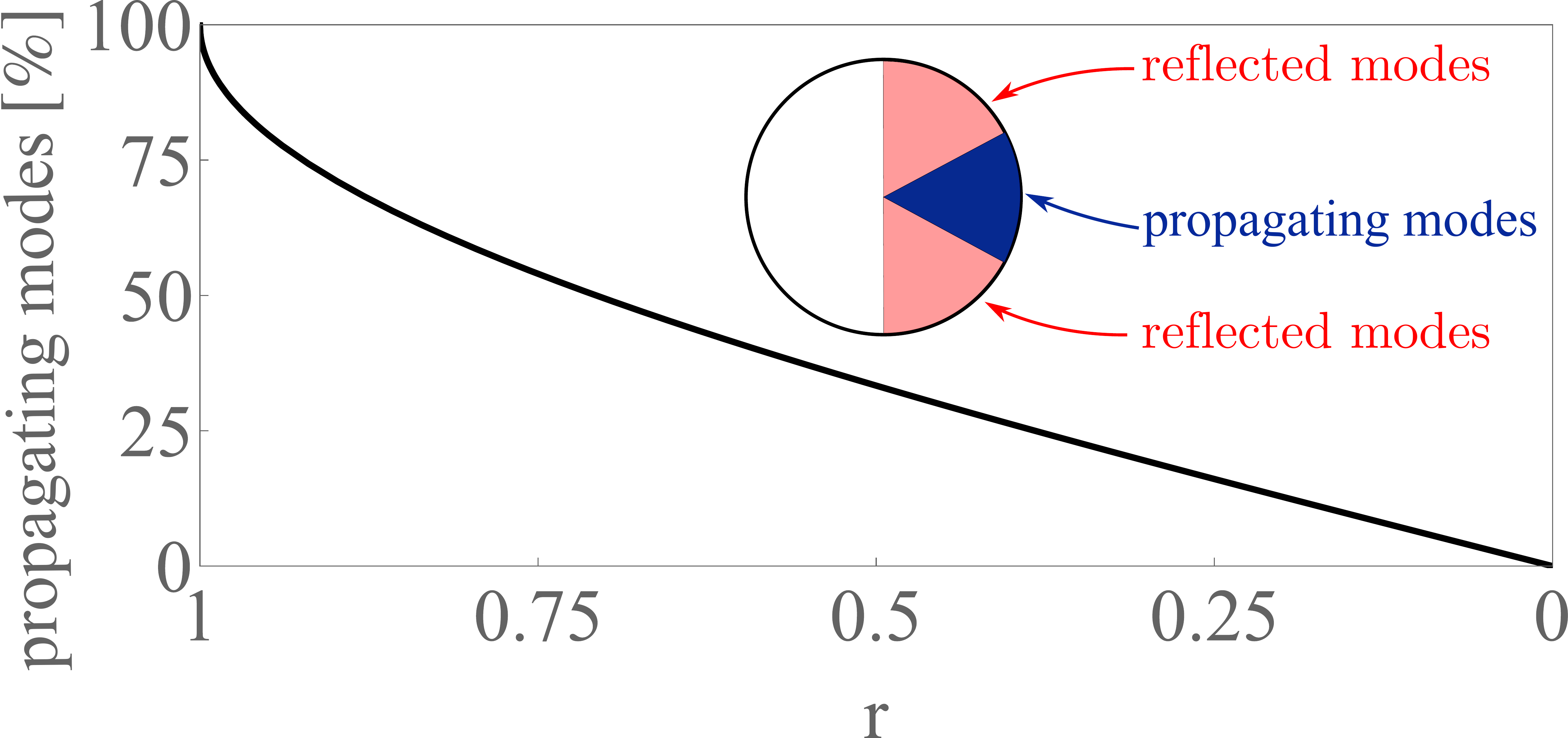}
	\caption{Fraction of the propagating modes, available for transmission,  as a function of FSM. The inset is drawn for a relatively small mismatch of $r=0.5$ for illustrative reasons.}
	\label{fig:3}
\end{figure}
Figure \ref{fig:3} illustrates the propagating modes as a function of $r$, which we define as the fraction of incident electrons from $L$ that have an angle of incidence smaller than the critical angle $\theta_c$. This fraction is given by $\theta_c/\pi=\arcsin(r)/\pi$. The other modes are reflected with unit probability due to the absence of corresponding modes in $R$. In a homogeneous setup ($r=1$), all modes can pass through the interface, but this fraction decreases gradually if $r$ is reduced from unity. After $r\approx0.5$, the number of available modes decays nearly linearly and vanishes as $r\to0$. For a practical superconductor-semiconductor system, $r$ reaches easily a value in the order of $0.01$, rendering the critical angle to a small cone.
\par
If we rescale $H\to Z\sqrt{\frac{\kappa_l \kappa_r}{m_lm_r}}\hbar^2$ and define $\tilde v=v_r/v_l$, we obtain the transmission coefficient for $E\ll\mu$ as
\begin{align}
	\label{eq:8}
	T(E,\theta_l)\approx T(\theta_l)=\frac{4\tilde v\;\Theta(\theta_c-\abs{\theta_l})}{\frac{4Z^2\tilde v}{\cos\theta_l\cos\theta_r}+\bk{1+\tilde v}^2}.
\end{align}
The step function expresses the mode selectivity, which is essentially determined by the FSM ratio $r$, and suppresses any transport across the junction for $\abs{\theta_l}>\theta_c$.
\par
\begin{figure}[htpb]
	\renewcommand{\figurename}{Fig.}
	\centering
	\includegraphics[width=.99\linewidth]{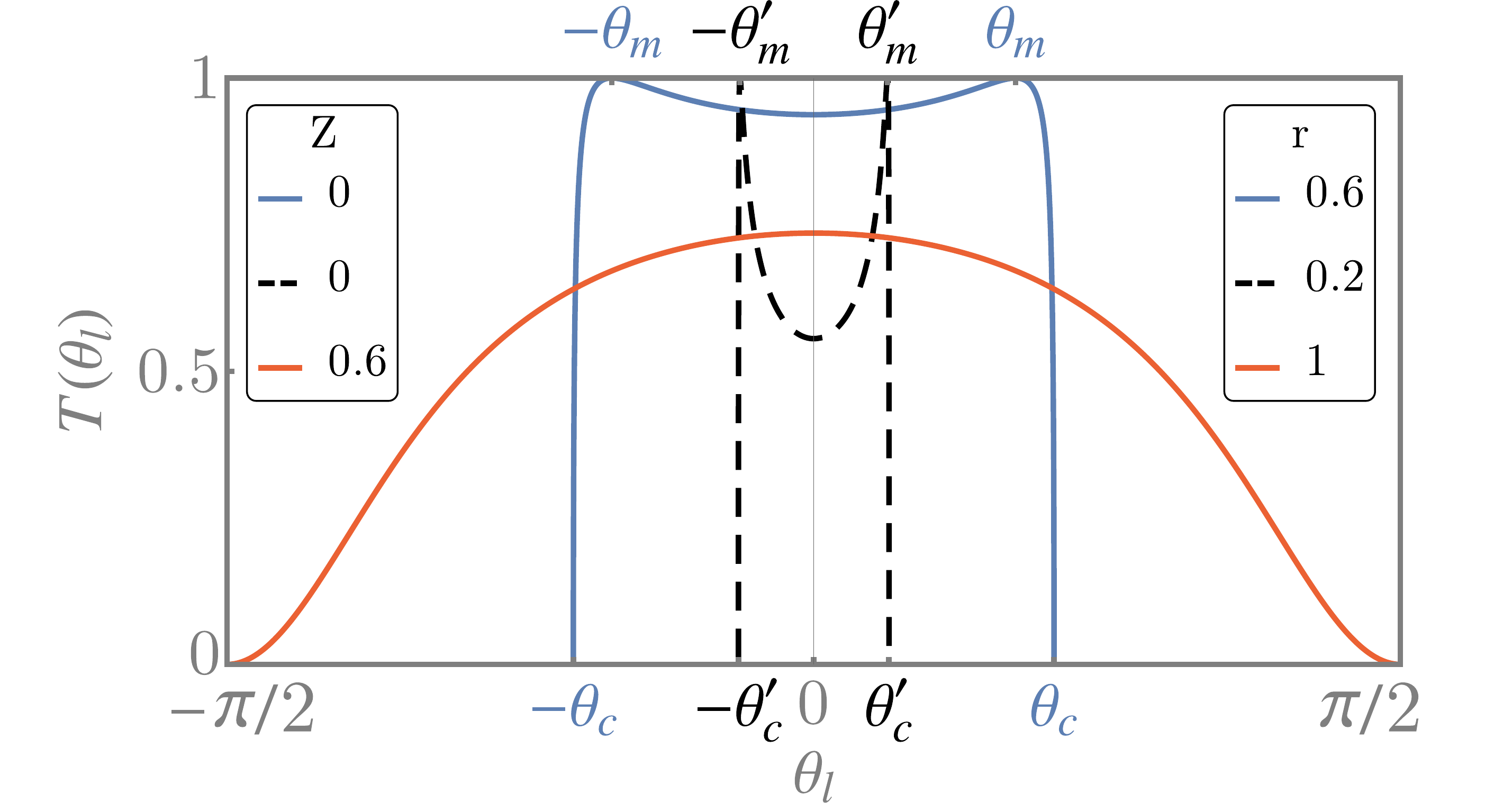}
	\caption{Transmission as a function of $\theta_l$ for different values of $Z$ and $r$.  The homogenous system with an elastic scattering potential (red) is compared to those with different degrees of electronic mismatch and $Z=0$ (blue, black). We choose $\mu=10^6\,\Delta_0$.}
	\label{fig:4}
\end{figure}

For perpendicular incidence, Eq. \eqref{eq:8} reduces to
\begin{align}
	\label{eq:9}
	T_0=T\bk{\theta_l=0}=\frac{4r}{4Z^2r+\bk{1+r}^2},
\end{align}
and we have $T_0<1$ for all $0<r<1$, even in the absence of an elastic scattering potential ($Z=0$). Eq. \eqref{eq:9} describes the 1D limit, where FSM  leads to non--perfect transmission due to the mismatch of the group velocities. Since it reduces the conductance in 1D junctions, FSM was earlier interpreted as an effective barrier, which further increased the repulsive potential at the interface \cite{Blonder83}. The implications of FSM are, however, richer in higher dimensions, allowing electrons with $\abs{\theta_l}<\theta_c=\arcsin\bk{\kappa_r/\kappa_l}$ to contribute strongly to the transmission, as illustrated in Fig. \ref{fig:4}.
\par
For the homogeneous case ($r=1$) and a finite $Z$ (red curve), the zero--mode conductance deviates from unity since the interface causes electron reflection. It decreases further for finite angles of incidence until it vanishes at $\theta_l=\pm\pi/2$.
\par
This behavior is distinctively different in a system featuring FSM for a clean interface (blue and black curves). First of all, we observe a sharp cut--off at $\theta_l=\pm\theta_c$, which is due to the absence of states in $R$ at larger angles of incidence. Moreover, the transmission coefficient \textit{increases} from the value for perpendicular incidence, $T_0$, when $\theta_l$ increases, and becomes unity at
\begin{align}
	\label{eq:10}
	\abs{\theta_l}=\theta_m\equiv\arcsin\bk{\frac{r}{\sqrt{1+r^2}}}<\theta_c.
\end{align}
This behavior can be attributed to the group velocities $v_l$ and $v_r$, which are, in general, different from each other if $r\neq1$. At $\theta_l=\pm\theta_m$, however, they coincide, allowing for perfect electron transmission across the interface. For larger angles of incidence, the transmission coefficient quickly decreases and vanishes for $\abs{\theta_l}\ge\theta_c$. Note that, for large FSM, $\theta_c$ approaches zero and nearly coincides with $\theta_m$, which yields sharp peaks at $\theta_l=\pm\theta_m\approx\pm\theta_c$ (black dashed line).
\par
\begin{figure}[htpb]
	\renewcommand{\figurename}{Fig.}
	\centering
    \includegraphics[width=.99\linewidth]{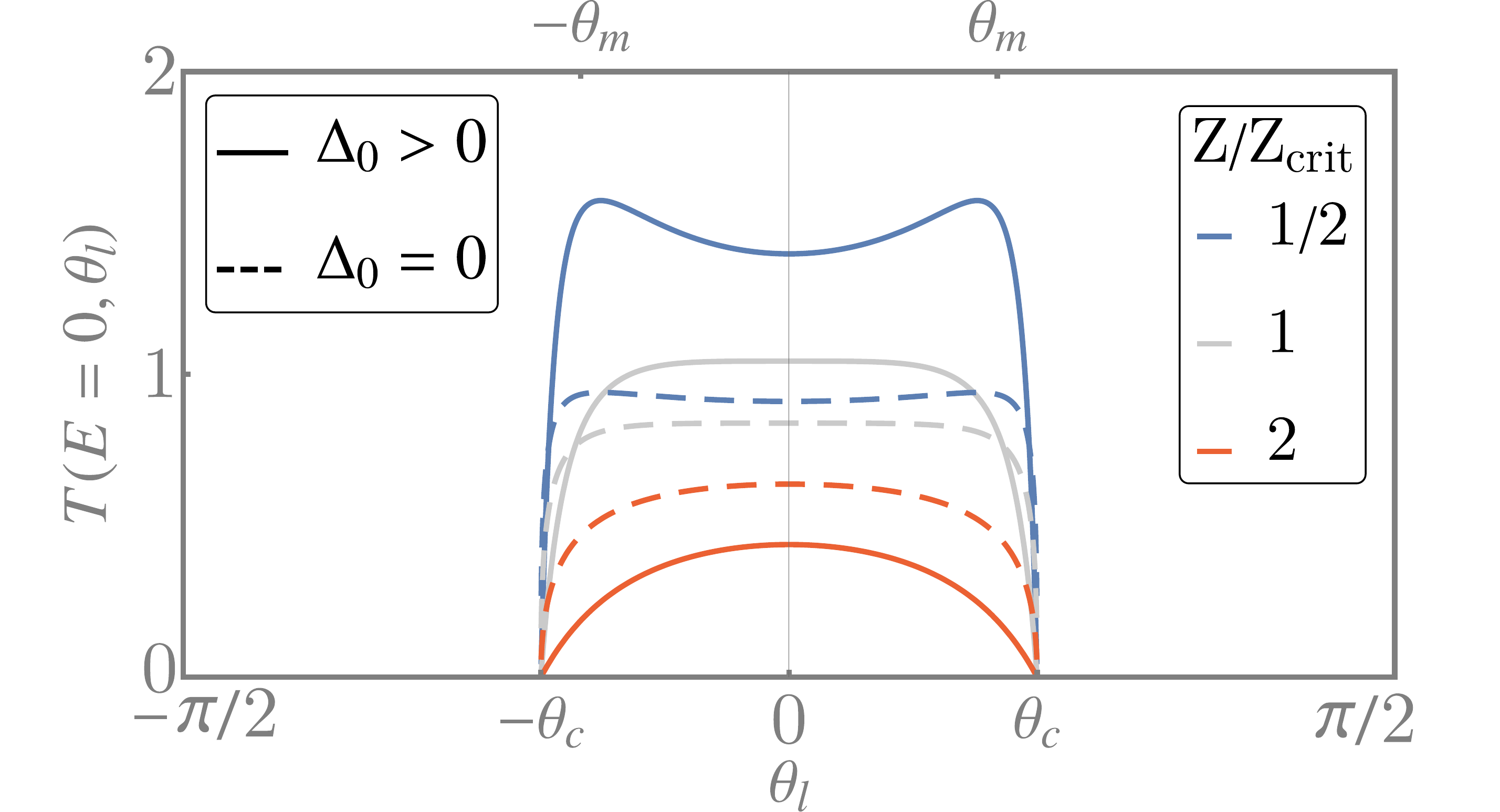}
	\caption{Transmission as a function of  $\theta_l$ for the NS-case (solid) and compared to the NN' case (dashed). We compare systems that are dominated (i) by FSM (blue), (ii) by the barrier (red) and (iii) those in the intermediate regime (gray). We choose $\mu=10^6\,\Delta_0$ and $r=0.6$.}
	\label{fig:5}
\end{figure}

In a case, in which both FSM and an interface barrier are present, a competition between the effects of the two quantities $r$ and $Z$ occurs. The dominant parameter can be identified by comparing $Z$ to a critical barrier strength defined by
\begin{align}
	\label{eq:11}
	Z_{\rm crit}\equiv\frac{\abs{1-r^2}}{2\sqrt{r(1+r^2)}}.
\end{align}
For $Z<Z_{\rm crit}$, we observe the same pattern as in a system with only FSM, while the transmission $T$ decreases monotonously from its zero--mode value if the elastic scattering potential at the interface dominates, $Z>Z_{\rm crit}$, see Fig. \ref{fig:5}.
Note that, even though FSM dominates for $Z<Z_{\rm crit}$, the maximum at $\theta_l=\pm\theta_m$ is no longer at unity.  A finite $Z$ promotes reflection at the interface, even if $v_l=v_r$.  The barrier--dominated transport, $Z>Z_{\rm crit}$, is affected in a similar manner by $r$, as the latter restricts the conductance to be finite only for $\abs{\theta_l}<\theta_c$, contrary to the homogeneous system where all incident modes, $\abs{\theta_l}<\pi/2$, have a certain probability to be transferred across the junction. Notably, FSM and interface barrier affect the transport rather differently.
\par
As shown in Fig. \ref{fig:5}, in which the metal is in the superconducting state, the angle-dependent features are also present, but the transmission is enhanced due to Andreev-reflection.  If $\Delta_0$ is finite, the transmission coefficient at the Fermi energy becomes
\begin{align}
	\label{eq:12}
	T(E=0,\theta_l)=\frac{8\tilde v^2\;\Theta(\theta_c-\abs{\theta_l})}{\bk{\frac{4Z^2\tilde v}{\cos\theta_l\cos\theta_r}+1+\tilde v^2}^2}.
\end{align}
In comparing this expression to Eq. \eqref{eq:8}, we find two essential differences: First, the maximum value of the conductance is twice that of the NN' setup, which is due to the additional hole channel in NS systems. Secondly, the barrier strength $Z$ enters the denominator in fourth order, while it is only second order in Eq. \eqref{eq:8}, which is a well-known characteristic of Andreev-reflection \cite{Blonder1982,Beenakker1992}. Therefore, the conductance is more sensitive to a scattering potential at the interface. Transmission across the junction in the normal state becomes thus more likely than Andreev reflection for $\Delta_0\neq 0$ and finite $Z$. We illustrate this behavior in Fig. \ref{fig:5} for $Z=2$. All this is a well-known subgap feature. For $E\gg\Delta_0$, however, the $Z$ sensitivity of the conductance is comparable in NN' and NS junctions.
 
\section{Conductance}
The given analysis of the selectivity of the interface is applicable from a metal into the semiconductor. A similar selectivity does not occur in the opposite direction, from a semiconductor to a metal. Therefore it is not possible to use the semiconductor and the metal as two separate equilibrium reservoirs to calculate the conductance. Instead we can consider a low carrier-density semiconductor sandwiched between two metal reservoirs, of which one can be superconducting. In analogy to the analysis of Octavio \textit{et al.} \cite{Octavio1983,Flensberg1988}, we distinguish inside the semiconductor, where we assume no energy--relaxation, right-moving and left-moving populations, $f_\rightarrow$ and $f_\leftarrow$, see App. \ref{app:C}. The equilibrium reservoirs are the two metallic electrodes attached to the semiconductor, which in the usual way take care of energy relaxation \cite{Nazarov2009}. In the commonly used geometric confinement, there are, for electrons in the reservoirs, \textit{closed} channels with $T=0$ and \textit{open} channels with $T\approx 1$. These open channels are characterized by the filling factors according to the Fermi-Dirac distribution of the reservoirs and by their directionality. 
\par
We proceed by calculating the current-voltage characteristic of the device shown in Fig. \ref{fig:1}. However, since we are interested in the energy-dependent properties of the conductance we take one of the electrodes as a normal metal, to avoid the complex dynamics of the Josephson-effect. This normal metal and the superconductor serve as equilibrium reservoirs in the spirit of Landauer-B\"uttiker-theory, whereas the semiconductor serves as the conductor in the central region. With this point of view, the metal-semiconductor-superconductor device will carry a current given by
\begin{widetext}
	\begin{align}
		\label{eq:13}
		I\bk{V}= \frac{1}{e\,R_N}  \int\limits_{-\theta_c}^{\theta_c} \D \theta_l \cos\theta_l\int\limits_{-\infty}^{\infty} \D E \bkk{f_0(E-eV)-f_0(E)}T(E,\theta_l),
	\end{align}
\end{widetext}
where $R_N$ is the normal--state resistance and $T(E,\theta_l)$ is defined by Eq. \eqref{eq:7}. For a symmetric arrangement like a superconductor-semiconductor-superconductor device, above the low voltage-range, the excess current-voltage characteristic will be given by Eq. \eqref{eq:13} multiplied by 2. For illustrative purposes, we provide explicit figures of $\partial I/\partial V$ and $I(V)$ in App. \ref{app:C}.


\section{Conclusion}
We have shown that, in the absence of other scattering processes, the conductance of a metal-semiconductor-metal device is limited by Fermi surface mismatch, leading to a directional selection of propagating modes in momentum space. We have applied this understanding to the electronic transport of superconductor-semiconductor-superconductor samples and argued that the observed excess current at applied voltages above the energy-gap can be understood based on this point of view. In addition, it has implications for our understanding of the proximity-effect under driven conditions \cite{Wiedenmann2017}. The tunability of the Fermi surface mismatch by a gate-voltage attached to the semiconductor provides a rich opportunity for further experimental and theoretical work.    
\begin{acknowledgments}
	We acknowledge inspiring discussions with the members of the experimental group of Martin Stehno and Laurens Molenkamp, which led to this study.  The work was supported by the DFG (SPP1666 and SFB1170 “ToCoTronics”), the Würzburg--Dresden Cluster of Excellence ct.qmat, EXC2147, project-id 390858490, and the Elitenetzwerk Bayern Graduate School on “Topological Insulators”.
\end{acknowledgments}
\appendix
\section{Model}
\label{app:A}
The eigensystem of the full kernel Hamiltonian, written in the basis $(\hat c,\;\hat c^\dagger)^T$,
\begin{align}
	{\mathcal H}_\mathrm{BdG}(x)=\mathcal H_0\tau_z+ \Delta(x)\tau_x
\end{align}
with
\begin{align}
\mathcal H_0=\hat k_x\frac{\hbar^2}{2m(x)}\hat k_x+\frac{\hbar^2k_y^2}{2m(x)}-\mu(x)+H\delta(x),
\end{align}
can be obtained by solving the Bogoliubov--de Gennes equation,
\begin{align}
	\label{eq:A2}
	{\mathcal H}_\mathrm{BdG}(x)\psi(x)=E\psi(x).
\end{align}
As introduced in the main text, the position--dependence of the effective masses as well as the superconducting order parameter and the electrochemical potential are given by 
\begin{subequations}
	\begin{align}
		m(x)&=m_l\Theta(-x)+m_r\Theta(x), \\
		\Delta(x)&=\Delta_0\Theta(x),\\
		\mu(x)&=\mu,
	\end{align}
\end{subequations}
which yields an interface at $x=0$ separating the left ($L$) from the right ($R$) domain. In the bulk regions, the eigenenergies are given by
\begin{subequations}
	\begin{align}
		E\big|_{x\ll 0}&=\frac{\hbar^2k^2}{2m_l}-\mu, \\
		E\big|_{x\gg 0}&=\sqrt{\bk{\frac{\hbar^2k^2}{2m_r}-\mu}^2+\Delta_0^2},
	\end{align}
\end{subequations}
with $k^2=k_x^2+k_y^2$. Furthermore, the eigenstates read
\begin{subequations}
	\begin{flalign}
		\psi^\pm_e(x)&= \begin{pmatrix}1 \\0\end{pmatrix}\E{\pm i k_e x}, &&
		\psi^\pm_h(x)= \begin{pmatrix}0 \\1\end{pmatrix}\E{\mp i k_h x}, \\
		\psi^\pm_{eq}(x)&= \begin{pmatrix}u \\v\end{pmatrix}\E{\pm i k_{eq} x}, &&
		\psi^\pm_{hq}(x)= \begin{pmatrix}v \\u\end{pmatrix}\E{\mp i k_{hq} x},
	\end{flalign}
\end{subequations}
where the subscripts and superscripts are explained in the main text and $u^2=1-v^2=\bk{1+\Omega/ E}/2$ with $\Omega=\sqrt{E^2-\Delta_0^2}$ are the superconducting coherence factors. The wave numbers read
\begin{subequations}
	\begin{align}
		\label{eq:A6}
		k_{e/h}&=\kappa_l\sqrt{1\pm\frac{E}{\mu}-\bk{\frac{k_y}{\kappa_l}}^2}, \\ k_{eq/hq}&=\kappa_r\sqrt{1\pm\frac{\Omega}{\mu}-\bk{\frac{k_y}{\kappa_r}}^2}
	\end{align}
\end{subequations}
with $\kappa_{l/r}=\sqrt{2m_{l/r}\mu}/\hbar$, and we obtain the group velocities according to the relation $v_g=\bk{\partial k/\partial E}^{-1}/\hbar$ as
\begin{align}
	v_{e/h}=\frac{\hbar k_{e/h}}{m_l},&&v_{eq/hq}=\frac{\hbar k_{eq/hq}}{m_r}\bk{\abs{u}^2-\abs{v}^2}.
\end{align}
The simplified relations in the main text are derived under the assumption $\mu\gg E,\Delta_0$. Then, we can drop the terms $E/\mu$ and $\Omega/\mu$ in Eq. \eqref{eq:A6} and obtain $k_e=k_h\equiv k_l$ and $k_{eq}=k_{hq}\equiv k_r$ as well as $v_e=v_h\equiv v_l$ and $v_{eq}=v_{hq}\equiv v_r$. 
\par
To determine the transport properties, we introduce the scattering state for an electron approaching the interface from the left asymptotic domain as
\begin{align}
	\phi(x)&=
	\begin{cases}
		\psi^+_e(x)+a\,\psi^-_h(x)+b\,\psi^-_e(x),& x<0\\
		c\,\psi^+_{eq}(x)+d\,\psi^+_{hq}(x),&x>0
	\end{cases}.
\end{align}
The coefficients are related to Andreev ($a$) and normal ($b$) reflection as well as transmission without ($c$) and with ($d$) branch crossing. The corresponding probability amplitudes read
\begin{subequations}
	\begin{align}
		A(E,k_y)&=\frac{v_h}{v_e}\abs{a_1}^2, && B(E,k_y)=\abs{b_1}^2,\\C(E,k_y)&=\frac{v_{eq}}{v_e}\abs{c_1}^2,&&D(E,k_y)=\frac{v_{hq}}{v_e}\abs{d_1}^2,
	\end{align}
\end{subequations}
and we obtain the transmission coefficient for electrical current by means of the relation  
\begin{align}
	T(E,k_y)=1+A(E,k_y)-B(E,k_y).
\end{align}
The scattering coefficients $a,b,c,d$ are determined by matching the waves appropriately at the interface, $x=0$, according to the conditions
\begin{subequations}
	\label{eq:A11}
	\begin{align}
		\lim\limits_{\varepsilon\to0}\bk{\phi_1(0+\varepsilon)-\phi_1(0-\varepsilon)}&=0,\\
		\lim\limits_{\varepsilon\to0}\bk{\frac{\phi_1'(0+\varepsilon)}{ m_r}-\frac{\phi_1'(0-\varepsilon)}{ m_l}}&=\frac{2H}{\hbar^2}\phi_1(0).
	\end{align}
\end{subequations}
\section{Transport properties}
\label{app:B}
Starting with the NN'-setup ($\Delta_0=0\,\Rightarrow\,A=D=0$), convincing ourselves that the wave numbers in $L$ can either be real or purely imaginary, and evaluating Eqs. \eqref{eq:A11}, this yields
\begin{flalign}
	T(E,k_y)&=1-B(E,k_y)=C(E,k_y)=&& \hfill \nonumber \\ &=\frac{4m_lm_r\hbar^4\mathrm{Re}(k_e)\mathrm{Re}(k_{eq})}{4H^2m_l^2m_r^2+\hbar^4\bk{m_rk_e+m_lk_{eq}}^2}.
\end{flalign}
Note that for $\Delta_0=0$, $k_{eq}$ describes an electron, not a quasiparticle. In order to obtain a finite conductance, both wave numbers (on the left and the right hand side of the junction) need to be real, imposing the condition $\bk{\abs{k_y}\le \kappa_l}\land \bk{\abs{k_y}\le\kappa_r}$, and we can simplify the transmission coefficient to
\begin{align}
	\label{eq:A13}
	T(E,k_y)=\frac{4\hbar^2v_ev_{eq}\Theta(\kappa_l-\abs{k_y})\Theta(\kappa_r-\abs{k_y})}{4H^2+\hbar^2\bk{v_e+v_{eq}}^2}.
\end{align}
Assuming $m_l>m_r$, a large electrochemical potential, $\mu\gg E$, and rescaling $H\to Z\sqrt{(\kappa_l \kappa_r)/(m_lm_r)}\hbar^2$, we obtain the NN' transmission coefficient $T(E,k_y)\approx T(k_y)$ as stated in Eq. \eqref{eq:8} of the main text,
\begin{align}
	T(k_y)=T(\theta_l)=\frac{4\tilde v\;\Theta(\theta_c-\abs{\theta_l})}{\frac{4Z^2\tilde v}{\cos\theta_l\cos\theta_r}+\bk{1+\tilde v}^2}.
\end{align}
\par 
The NS--setup is generally more complicated, in particular due to the additional hole channels. However, under the assumption $\mu\gg E,\Delta_0$, we obtain the analytic expression
\begin{widetext}
	\begin{align}
		T(E, k_y)=	\begin{cases}
			\frac{8\tilde v^2\,\Theta(\kappa_l-\abs{k_y})\Theta(\kappa_r-\abs{k_y})}{\bk{\frac{4Z^2\tilde v}{\cos\theta_l\cos\theta_r}+1+\tilde v^2}^2-\bk{\frac{E}{\Delta_0}}^2\bk{\frac{16Z^4\tilde v^2}{\cos^2\theta_l\cos^2\theta_r}+\frac{8Z^2\tilde v\bk{1+\tilde v^2}}{\cos\theta_l\cos\theta_r}+\bk{1-\tilde v^2}^2}}, & E<\Delta \\
			\frac{4\tilde v\,\Theta(\kappa_l-\abs{k_y})\Theta(\kappa_r-\abs{k_y})}{\bk{\frac{4Z^2\tilde v}{\cos\theta_l\cos\theta_r}+1+\tilde v^2}\bk{u^2-v^2}+2\tilde v}, & E>\Delta
		\end{cases}.
	\end{align}
\end{widetext}
This reduces to Eq. \eqref{eq:12} in the main text for $E=0$ and to the normal--state transmission, Eq. \eqref{eq:8}, for $E\gg\Delta_0$.
\par 
The competition between FSM and interface barrier is moderated by the quantity $Z_\mathrm{crit}$, as we introduce it in the main text. For a strong repulsive barrier, $Z>Z_\mathrm{crit}$, the conductance decreases monotonously from its zero--mode value, while it increases from $T_0$ and reaches a maximum at $\abs{\theta_l}=\theta_m$ if $Z<Z_\mathrm{crit}$.  A rigorous, but tedious method to obtain an explicit expression for this critical barrier strength is to demand $\theta_m$ to be real. While this is always true for $Z=0$, a finite interface potential $Z>Z_\mathrm{crit}$ shifts the position of the maxima into the complex domain. Conveniently, a second order Taylor expansion of Eq. \eqref{eq:A13}, for $\mu\gg E,\Delta_0$ and $m_r=r^2m_l$, around $k_y=0$,
\begin{align}
	T(k_y\approx0)=T_0 + T_2\frac{k_y^2}{\kappa_l^2}+\mathcal O(k_y^4),
\end{align}
with 
\begin{subequations}
	\begin{align}
		T_0&=\frac{4r}{4rZ^2+\bk{1+r}^2},\\
		T_2&=\frac{2\bk{1-r^2}^2-8rZ^2\bk{1+r^2}}{r\bk{4rZ^2+\bk{1+r}^2}},
	\end{align}
\end{subequations}
yields exactly the same result as the more complicated method mentioned above. By demanding $T_2$ to be positive, such that $T$ increases from its zero--mode value $T_0$, we find
\begin{align}
	\label{eq:A17}
	Z<Z_\mathrm{crit}=\frac{\abs{1-r^2}}{2\sqrt{r\bk{1+r^2}}}.
\end{align}
This allows us to identify whether FSM or $Z$ dictates the mode--dependence of the conductance.
\section{Derivation of the current}
\label{app:C}
From the transmission, we directly obtain the currents in NN' and NS systems, and thus the excess current. To motivate Eq. \eqref{eq:13} in the main text, we start from the single--mode expression for the current density
\begin{align}
	J(k_y)=2e\int\limits_{-\infty}^\infty N_{k_y}(E)v_{k_y}(E)\bkk{f_{k_y,\rightarrow}(E)-f_{k_y,\leftarrow}(E)}\D E
\end{align}
with $N_{k_y}(E)$ and $v_{k_y}(E)$ the density of states and group velocity for each 1D mode $k_y$ at energy $E$, respectively. We can further simplify this equation by means of the relation $N_{k_y}(E)=1/\bkk{2\pi\hbar v_{k_y}(E)}$ and obtain
\begin{align}
	J(k_y)=\frac{2 e}{h}\int\limits_{-\infty}^\infty \bkk{f_{k_y,\rightarrow}(E)-f_{k_y,\leftarrow}(E)}\D E.
\end{align}
Assuming equilibrium reservoirs connected to $L$ and $R$, the distribution functions are given by

\begin{subequations}
	\begin{align}
			f_{k_y,\rightarrow}(E)=&f_0(E-eV), \\
			f_{k_y,\leftarrow}(E)=&A(E,k_y)\bkk{1-f_{k_y,\rightarrow}(-E)}\nonumber \\&+B(E,k_y)f_{k_y,\rightarrow}(E)\nonumber \\&+\bkk{C(E,k_y)+D(E,k_y)}f_0(E)
	\end{align}
\end{subequations}
and we obtain
\begin{align}
	J(k_y)=\frac{2 e}{h}\int\limits_{-\infty}^\infty \bkk{f_{0}(E-eV)-f_{0}(E)}T(E,k_y)\D E.
\end{align}
Considering FSM and the contributions from all modes, we obtain the current for $\mu\gg E,\Delta_0$ according to Eq. \eqref{eq:13} in the main text.\par
\begin{figure}[htpb]
	\renewcommand{\figurename}{Fig.}
	\centering
	\includegraphics[width=.79\linewidth]{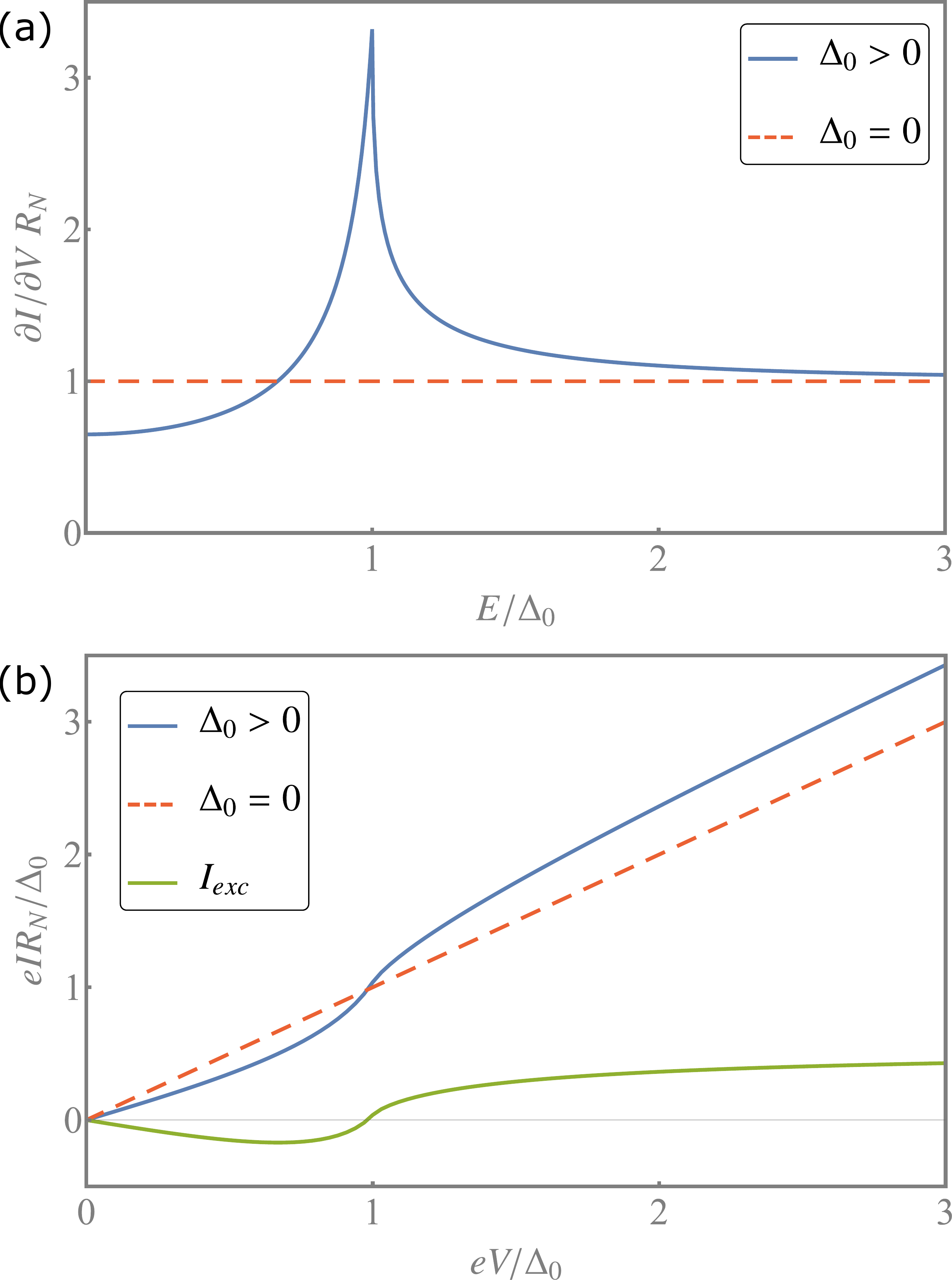}
	\caption{(a) $\partial I/\partial V$ as a function of the energy $E$ and (b) current as a function of the bias $eV$ for $r=0.2$ and $Z=Z_{crit}/4$, summed over all modes. The offset between $I(V)\big|_{\Delta_0>0}$ and $I(V)\big|_{\Delta_0=0}$ is denoted the excess current $I_{exc}(V)$ and reaches its asymptotic value for $eV\gg\Delta_0$. }
	\label{fig:A1}
\end{figure}

  
To conclude, we plot the differential conductance as well as the current according to Eq. \eqref{eq:13} for a particular FSM $r=0.2$ and a weak barrier $Z=Z_\mathrm{crit}/4$ in Fig. \ref{fig:A1}. 
While the normal--state transmission is constant, $\frac{\partial I}{\partial V}\big|_{\Delta_0>0}$ features a resonance at $E=\Delta_0$ before it converges to $\frac{\partial I}{\partial V}\big|_{\Delta_0=0}$ for $E>\Delta_0$. This results in an offset between the currents $I(V)\big|_{\Delta_0>0}$ and $I(V)\big|_{\Delta_0=0}$ at large biases, while their slope is equal. This offset defines the excess current, which is, more generally, given by
\begin{align}
	I_\mathrm{exc}(V)=I(V)\big|_{\Delta_0>0}-I(V)\big|_{\Delta_0=0}.
\end{align}
As we can see, this excess current is finite even for a notable FSM, as long as the interface barrier remains small, i.e., $Z<Z_\mathrm{crit}$. 
\par 
These results have been obtained under the assumption of equilibrium reservoirs on each side of the semi\-conductor--superconductor interface. In practice, in the application to the experimental results such as for example to the S-HgTe-S system,under bias conditions thermalisation in HgTe is negligible. Therefore, a more detailed description, in the spirit of the Octavio--Tinkham--Blonder--Klapwijk theory \cite{Octavio1983}, is needed (in preparation).

\footnotetext[1]{For simplicity, we assume a step--wise variation of parameters across the junction, which should be justified if the Fermi wave lengths of the two regions (on the two sides of the junction) differ substantially.}
\footnotetext[3]{The factor $2$ in Eq. \eqref{eq:1} takes spin degeneracy into account.}
\bibliographystyle{apsrev4-2}

\end{document}